\documentclass[reqno,12pt]{amsart}
        \newtheorem{theorem}{Theorem}[section]
        \newtheorem{theorem*}[theorem]{Theorem}
        \newtheorem{lemma}[theorem]{Lemma}
        \newtheorem{definition}[theorem]{Definition}
        
        \newtheorem{proposition}[theorem]{Proposition}
        \newtheorem{corollary}[theorem]{Corollary}
        \newtheorem{example}{Example}
        \theoremstyle{remark}
        \newtheorem{remark}[theorem]{Remark}
        \numberwithin{equation}{section}

        \usepackage{array}
        
        \newcolumntype{P}[1]{>{\centering\arraybackslash}p{#1}}
        \usepackage{makecell}
        \usepackage{float}
        
        \usepackage{multirow}
        \usepackage{array,booktabs}
        \usepackage{todonotes}
        \usepackage[margin=1in]{geometry}
        \usepackage{verbatim}
        \usepackage{amsmath}
        \usepackage{amsthm}
        \usepackage{amsfonts}
        \usepackage{amssymb}
        \usepackage{amsthm}
        \usepackage{enumerate}
        \usepackage{amsrefs}
        \usepackage{enumerate}
        \usepackage{amsrefs}
        \usepackage{tensor}
        \usepackage{epic}
        \usepackage{mathtools}
        \usepackage{tikz}
        \usetikzlibrary{chains}
        \usepackage{graphicx}
        \usepackage[font=small,skip=0pt]{caption}
        \usepackage{makecell}
        \usepackage{tensor}
        \usepackage[super]{nth}
        \usepackage{tabulary}
        \usepackage{lipsum}
        \usepackage{cases}
        \usepackage{accents}

        \AtEndDocument{\bigskip{\footnotesize%
        (M. Alkadhem) \textsc{Department of Mathematics, College of Science and Humanities, Imam Abdulrahman Bin Faisal University, Jubail 35811, Saudi Arabia.} 
          \textit{E-mail address} \,\texttt{malkadhem@iau.edu.sa} {ORCID: 0000-0001-6815-4135}
           \par
           \addvspace{\medskipamount}
          (M. Feigin) \textsc{School of Mathematics and Statistics, University of Glasgow, Glasgow, United Kingdom of Great Britain and Northern Ireland.} 
          \textit{E-mail address} \, \texttt{misha.feigin@glasgow.ac.uk} 
             ORCID: 0000-0001-9184-3940 }}
        
        \title{Oriented V-systems}
        \author{Maali Alkadhem and Misha Feigin }

\begin{document}

        \maketitle

\begin{abstract}
We introduce the notion of an oriented $\vee$-system which gives a class of solutions of the oriented associativity equations generalizing Veselov's $\vee$-system solutions of WDVV equations.
An oriented $\vee$-system is a special configuration of pairs of a vector and a covector in a linear space. 
We show that the class of oriented $\vee$-systems is closed under taking subsystems and restrictions. We provide examples which include a multiparameter family related to imprimitive complex reflection groups.

\end{abstract}

        \section{Introduction}

The associativity equations, also known as the WDVV equations, play a fundamental role in several areas of mathematical and theoretical physics, and geometry. They have their origins in the study of two-dimensional topological field theory at the end of 1980’s by Witten, Dijkgraaf, E. Verlinde and H. Verlinde
\cite{Witten1990,DVV1990}. 
These are the nonlinear third order partial differential equations for a single scalar function $\mathcal{F}(x)$, $x=(x_1,\dots ,x_N)$, called the prepotential, and they express the associativity condition for a family of commutative algebras parametrised by $x$.

The WDVV equations have the form
\begin{equation}\label{WDVV.1}
\frac{\partial^3 \mathcal F}{\partial x_i\partial x_j\partial x_m}\eta^{mn}
\frac{\partial^3 \mathcal F}{\partial x_n\partial x_k\partial x_s}
=
\frac{\partial^3 \mathcal F}{\partial x_i\partial x_k\partial x_m}\eta^{mn}
\frac{\partial^3 \mathcal F}{\partial x_n\partial x_j\partial x_s},
\end{equation}
where $\eta^{mn}$ is a constant non-degenerate symmetric matrix which plays a role of a (contravariant) flat metric.
The third order derivatives of the function $\mathcal F$ determine both $\eta$ and the structure constants of a family of commutative associative algebras. 
These equations were given a geometric interpretation in 
Dubrovin's theory of Frobenius manifolds \cite{Dubrovin1996}. 
The WDVV equations play a fundamental role in enumerative geometry and quantum cohomology as the generating functions of genus zero Gromov--Witten invariants satisfy these equations \cite{Maninbook}.

There is a remarkable class of solutions of the WDVV equations of the form
\begin{equation}
\label{Fcheck}
\mathcal F(x)=\sum_{\alpha\in \mathcal{A}} \alpha(x)^2 \log \alpha(x),
\end{equation}
where $\mathcal{A}$ is a finite collection of covectors on a vector space $V\cong \mathbb{C}^N$. For an arbitrary root system these solutions were given in \cites{MartiniGragert1999, Veselov1999}, and they also describe almost dual Frobenius structure on the Coxeter orbit space Frobenius manifolds \cite{Dubrovin2004}.
In the case of classical root systems such solutions appear also in \cite{MMM2000} as leading terms of Seiberg--Witten prepotentials. 
Furthermore, Veselov initiated the study of solutions of the form \eqref{Fcheck} for a general configuration of covectors $\mathcal{A}$, and he reformulated such configurations in a linear algebraic way which gave the notion of a $\vee$-system \cite{Veselov1999}. 

The class of $\vee$-systems was extensively studied. In particular, it was shown that this class is closed under the natural operations of taking subsystems and restrictions \cites{MishaVeselov2007,MishaVeselov2008}.
 The latter is related to Dubrovin's duality on the discriminant strata of Coxeter orbit space Frobenius manifolds \cite{MishaVeselov2007}, \cite{AFS}. 
A classification of $\vee$-systems with up to $27$ covectors in a real three-dimensional space, which is the first dimension where $\vee$-conditions are non-trivial, has been recently obtained 
in \cite{BPV.2026}. In all the known cases (see \cite{MishaVeselov2008}) the underlying configurations of hyperplane arrangements  are Coxeter restricted arrangements.  A close relation of theories of $\vee$-systems and polyhedral K\"ahler metrics in a projective space \cite{dBP2025} was established in \cite{BPV.2026}. 


There is a generalization of WDVV equations known as {\it orineted associativity equations} following Losev and Manin \cite{LM2004}. In this case the scalar prepotential $\mathcal F$ is replaced by a vector potential $f = (f^1,\dots, f^N)$. Geometrically this comes from a non-symmetric treatment of marked points on the rational curves whose maps to the target define quantum cohomology of the space, which justifies the terminology.
The corresponding partial differential equations have order two and they are of the form
\begin{equation}\label{O.A.Q.1}
\frac{\partial^2 f^i}{\partial x_j\partial x_m}
\frac{\partial^2 f^m}{\partial x_k\partial x_n}
=
\frac{\partial^2 f^i}{\partial x_k\partial x_m}
\frac{\partial^2 f^m}{\partial x_j\partial x_n}.
\end{equation}
This system can be considered as a natural generalization of the WDVV equations. 
Indeed, 
given a prepotential $\mathcal F$ satisfying \eqref{WDVV.1}, the vector potential  
\begin{equation}
f^i=\eta^{im}\frac{\partial \mathcal F}{\partial x_m}
\end{equation}
gives a solution of \eqref{O.A.Q.1}.

It is explained by Manin in \cite{Manin2005} that oriented associativity equations \eqref{O.A.Q.1} imply that one has the structure of an $F$-manifold \cite{He} for the underlying multiplication which can be defined in terms of $f$.  Conversely, equations \eqref{O.A.Q.1} mean that the $F$-manifold has an additional compatible flat structure \cite{Manin2005}. 
These flat $F$-manifolds were studied by Cotti who described their germs precisely in terms of the monodromy data in the homogeneous semisimple case, and also investigated convergence of formal vector potentials \cite{Cotti2024}.
An approach to solutions of the oriented associativity equations through a generalization of cohomological field theory was developed in \cite{BR}.
The oriented associativity equations were also studied 
in relation with conjugate curvilinear coordinate nets and hydrodynamic type equations \cite{PavlovSergyeyev2014}. 


A specialization of oriented associativity equations arises in the open Gromov--Witten theory. The open version of the WDVV equations appeared in the work of Horev and Solomon \cite{HS12}.
In the open theory one has two prepotentials $F$ and $F^o$.
The function $F$ satisfies the ordinary WDVV equations and $F^o$ satisfies the second order differential equations once a solution $F$ is fixed. 
This is a particular case of oriented associativity equations when vector  potential can be expressed through two functions $F$ and $F^o$.
Recent work of Proserpio and Strachan extends the
\(\vee\)-systems type ansatz to the solutions of open WDVV equations, deriving linear 
algebraic conditions on corresponding covector configurations in the open theory \cite{ProserpioStrachan2026}  (see also \cite{B&B19}, \cite{Alm} for some further developments on the open WDVV equations). 

In the present paper, we are interested in the general case of oriented associativity equations and a particular class of its solutions which generalizes solutions \eqref{Fcheck} of WDVV equations. 
More specifically, 
the vector potential has the form
\[
f(x)
=
\sum_{\alpha\in\mathcal A}
\alpha(x)\log\alpha(x)\,a_\alpha,
\qquad
a_\alpha\in V.
\]
The oriented associativity equations express associativity of a family of algebras with the product
\begin{equation}
    \label{mult}
u*v = \sum_{\alpha \in \mathcal A} \frac{\alpha(u)\alpha(v)}{\alpha(x)} a_\alpha,
\end{equation}
where $u, v \in V$ and $x\in V$ satisfies $\alpha(x)\ne 0$ for all $\alpha \in \mathcal A$.

In contrast to the ordinary $\vee$-system, where the covectors are identified with associated vectors through a non-degenerate bilinear form, the present
construction allows the assignment
\[
\Phi:\mathcal A\longrightarrow V,
\qquad
\alpha\longmapsto a_\alpha,
\]
to be independent of a metric which may not exist. This makes the ansatz naturally adapted to the metric-free settings. 

The ordinary $\vee$-systems can also be formulated as a flatness condition of Dubrovin type connection \cite{Veselov1999}. They also come together with a natural non-degenrate bilinear form on $V$ defined directly in terms of covectors $\mathcal A$. Another class of flat connections of so-called Dunkl--Kohno type with singularities along a configuration of hyperplanes was studied by Couwenberg, Heckman and Looijenga in \cite{CHL.2005}. In this case, one has a non-degenerate Hermitian form which does not have to be present in the context of $\vee$-systems. These connections arise in the theory of bi-flat $F$-manifolds by Arsie and Lorenzoni \cite{ArsieLorenzoni2017} in relation to (well-generated) complex reflection groups orbit spaces and an extension of Dubrovin's almost duality to these settings. Further results on a generalization of almost duality for orbit spaces of complex reflection groups were obtained by Konishi, Minabe and Shiraishi in 
\cite{KMS}, \cite{KM} (see also \cite{KatoMS1}, \cite{KatoMS2}).

A motivation behind our work is to extend the theory of $\vee$-systems so that to cover all these flat connections without extra constraints such as the existence of a Hermitian metric. In fact, connections of such more general type were considered by de Borbon and Panov in \cite{dBP1}, \cite{dBP2}. In the context of $F$-manifolds we deal with an additional flat structure compatible with  multiplication as defined in \cite{Manin2005}. 

We reformulate the oriented associativity equations for $f(x)$ as linear algebraic conditions on the collection of pairs $(\alpha,a_\alpha)\in V^*\times V$. 
This leads to the notion of an
\textit{oriented $\vee$-system}. In contrast to the ordinary WDVV case the oriented $\vee$-system conditions are non-trivial in dimension two and the corresponding configurations can be described. 

We further study these systems. 
We introduce subsystem and restriction procedures for arbitrary oriented $\vee$-systems which produce new oriented
\(\vee\)-systems starting with a given one.  
We also relate the corresponding associative 
multiplication \eqref{mult} to a pencil of flat connections. This leads to an $F$-manifold with the multiplication compatible with both flat structure and a flat identity vector field, also known as a flat $F$-manifold. 

A rich source of examples comes from finite complex reflection groups. In particular, the restriction procedure applied to the infinite family of imprimitive complex reflection groups $G(m, 1, N)$ leads to a cyclotomic generalization of multiparameter Lauricella type $\vee$-systems from \cite{ChalykhVeselov2001}. 
        
        \section{Oriented associativity equations}\label{section.o.a.e. and v-system}
        Let $V\cong \mathbb{C}^N, N\geq 2$ with the standard basis $e_1,\dots, e_N$, and let $x_1,  \dots, x_N$ be the corresponding coordinates in $V$.
Let $\partial_i =\frac{\partial}{\partial x_i}$ for $1 \leq i\leq N$.
        Consider the bilinear multiplication  
        $$\ast \colon T_x V \times T_x V \to T_x V,$$
     where $x\in V$, defined on the basis of $T_x V$ by
        \begin{equation}\label{product}
\partial_i \ast \partial_j \coloneqq \sum_{k=1}^{N} C_{ij}^{k}\partial_k.
\end{equation}

Consider a vector-valued function $f\colon V\longrightarrow V,$
$$ f(x)=(f^1 (x),\dots , f^N (x)),$$
where $f^i\colon V\to \mathbb{C}$ are some functions and $x=(x_1, \dots , x_N).$
Let us assume that the structure constants of the multiplication \eqref{product} have the following form:
\begin{equation}\label{matrix entries.1}
C^k_{ij}
=
\frac{\partial^2f^k}
{\partial x_i\partial x_j},
\qquad
1\leq i,j,k\leq N.  
\end{equation}
Note that the multiplication $\ast$ is then commutative.
Let us also define the $N\times N$ matrices $C_p, 1\leq p\leq N$ with the matrix entries 
\begin{equation}
\label{Cp-matrix}
(C_p)^{i}_{j}
=
\frac{\partial^2 f^i}
{\partial x_p\partial x_j},
\qquad
1\leq i,j\leq N.  
\end{equation}
\begin{proposition} \label{asossiative=oriented}
The associativity of multiplication \eqref{product} is equivalent to the commutativity equations 
\begin{equation}
    \label{O.A.E}
[C_p , C_r]=0, \quad \forall \, p,r=1,...,N.
\end{equation}
\end{proposition}
\begin{proof}
The associativity of the multiplication \eqref{product} means that $\forall \, p,q,r$
 \begin{align*}
  \partial_p \ast (\partial_q \ast \partial_r) &=( \partial_p \ast \partial_q) \ast \partial_r.   
 \end{align*}
  Equivalently, $
     C^m_{pt} C^t_{qr}  =   C^m_{sr} C^s_{pq}$, which is the commutativity relation $[C_p, C_r]=0$.   
\end{proof}
 Following \cite{LM2004} we will refer to the system of equations \eqref{Cp-matrix}, 
\eqref{O.A.E} for a function $f(x)$ 
as the \textit{oriented associativity equations}.

Using the canonical isomorphism $T_xV\cong V$, we can also regard $f$
as the vector field
$$f=\sum_{k=1}^N f^k(x)\partial_k.$$
    The multiplication \eqref{product} can equivalently be written as
    \begin{equation*}
        \partial_i \ast \partial_j=[\partial_i, [\partial_j,f]].
    \end{equation*}
Following \cite{Manin2005}
we refer to $f$ as the \textit{vector potential} for the multiplication \eqref{product} with the structure constants \eqref{matrix entries.1}.
   
   The oriented associativity equations are invariant under invertible
linear changes of coordinates, provided that the vector potential $f$
is transformed accordingly as a vector field.

\section{Oriented $\vee$-systems}

Let $\mathcal A\subset V^*$ be a finite set of non-zero covectors,
and let $\Phi\colon\mathcal A\longrightarrow V$
be a map. For each $\alpha\in\mathcal A$, set
$a_\alpha=\Phi(\alpha).$
We are interested in the logarithmic vector potentials of the form
\begin{equation}\label{logarithmic solution}
f(x)= f_{\mathcal{A},\Phi}(x)
=
\sum_{\alpha\in\mathcal A}
\alpha(x)\log\alpha(x)\,a_\alpha.
\end{equation}
Such logarithmic potentials can be understood locally or as multivalued functions. In any case their second derivatives and the associated structures are well defined on the arrangement complement. The corresponding associative product is given by formula \eqref{mult}. 

Let us describe two elementary operations on pairs
$(\mathcal A,\Phi)$ such that the validity of the corresponding oriented
associativity equations is preserved.

First, suppose that
\[
\beta_1=c_1\beta,
\qquad
\beta_2=c_2\beta,
\]
are two collinear covectors in $\mathcal A_1$, where
$c_1,c_2\in\mathbb C^\times$ and $\beta\in V^*\setminus\{0\}$.
We replace $\beta_1,\beta_2$ by the single covector $\beta$ and define
\[
\Phi_2(\beta)
=
c_1\Phi_1(\beta_1)
+
c_2\Phi_1(\beta_2).
\]
For every other $\gamma\in\mathcal A_1$, we retain
\[
\gamma\in\mathcal A_2,
\qquad
\Phi_2(\gamma)=\Phi_1(\gamma).
\]

The oriented associativity equations for the vector potential \eqref{logarithmic solution} for $(\mathcal{A}_1, \Phi_1)$ and $(\mathcal{A}_2, \Phi_2)$ are equivalent since the corresponding  vector potentials differ by a linear function. 

Conversely, a single covector $\beta$ may be replaced by two
collinear covectors
\[
\beta_1=c_1\beta,
\qquad
\beta_2=c_2\beta,
\]
provided that
\[
c_1\Phi_2(\beta_1)
+
c_2\Phi_2(\beta_2)
=
\Phi_1(\beta).
\]

Second, a covector $\beta$ with 
$\Phi(\beta)=0$
may be added to or removed from the configuration.

We say that two pairs $(\mathcal A_1,\Phi_1)$ and
$(\mathcal A_2,\Phi_2)$ are \textit{equivalent} if one can be obtained from the
other by a finite sequence of these elementary operations.

Consequently, without loss of generality, we will be assuming  further in this section that the covectors in $\mathcal A$ are pairwise
non-collinear and
$a_\alpha=\Phi(\alpha)\neq 0$ for every $\alpha\in\mathcal A.$

The following lemma holds.
\begin{lemma}\label{identity on L}
The oriented associativity equations \eqref{Cp-matrix}, \eqref{O.A.E} for the vector potential \eqref{logarithmic solution} are equivalent to the identity 
  \begin{equation}\label{main identity}
         \sum_{\beta \in \mathcal{A}} \Big( \frac{ \gamma (a_\beta) \beta \otimes a_{\gamma} - \beta (a_{\gamma}) \gamma \otimes a_{\beta}  }{ \beta (x)  } \Big)\gamma \wedge \beta   =0    
    \end{equation}
    on the hyperplane $\gamma (x)=0$ for all $\gamma \in \mathcal{A}$.  
\end{lemma}

\begin{proof}
For $\alpha\in\mathcal A$, write $\alpha_i=\alpha(e_i),$
and define the map $\alpha\otimes a_{\alpha}\colon V\to V$ by
$$(\alpha\otimes a_\alpha)(v)
=
\alpha(v)a_\alpha,
\qquad v\in V.$$
The matrix $C_i$ is the matrix of the linear operator
\[
\mathcal C_i
=
\sum_{\alpha\in\mathcal A}
\frac{\alpha_i}{\alpha(x)}
\,\alpha\otimes a_\alpha
\]
since
\[
\begin{aligned}
\mathcal C_i(e_k)
&=
\sum_{\alpha\in\mathcal A}
\frac{\alpha_i\alpha_k}{\alpha(x)}
a_\alpha.
\end{aligned}
\]
%
Therefore, the oriented associativity equations 
$[\mathcal C_i,\mathcal C_j]=0$ are equivalent to
\begin{equation}\label{equivalent to OAE}
\sum_{\alpha,\beta\in\mathcal A}
\frac{\alpha_i\beta_j}
{\alpha(x)\beta(x)}
\Bigl(
\alpha(a_\beta)\beta\otimes a_\alpha
-
\beta(a_\alpha)\alpha\otimes a_\beta
\Bigr)=0  
\end{equation}
for all $i,j$.

Since the covectors in $\mathcal A$ are pairwise
non-collinear, all the possible poles in the left-hand side of \eqref{equivalent to OAE}
are simple and lie on the
hyperplanes $\alpha(x)=0$.
Fix $\gamma\in\mathcal A$. The residue along the hyperplane
$\gamma(x)=0$ is obtained from the terms for which either
$\alpha=\gamma$ or $\beta=\gamma$. 
Thus, the residue at $\gamma(x)=0$ is
\[
\begin{aligned}
\sum_{\beta\neq\gamma}
\frac{\gamma_i\beta_j}{\beta(x)}
\Bigl(
\gamma(a_\beta)\beta\otimes a_\gamma
-
\beta(a_\gamma)\gamma\otimes a_\beta
\Bigr)+
\sum_{\alpha\neq\gamma}
\frac{\alpha_i\gamma_j}{\alpha(x)}
\Bigl(
\alpha(a_\gamma)\gamma\otimes a_\alpha
-
\gamma(a_\alpha)\alpha\otimes a_\gamma
\Bigr).
\end{aligned}
\]
Renaming $\alpha$ as $\beta$ in the second sum, 
the residues take the form
\[
\begin{aligned}
\sum_{\beta\neq\gamma}
\frac{\gamma_i\beta_j-\beta_i\gamma_j}{\beta(x)}
\Bigl(
\gamma(a_\beta)\beta\otimes a_\gamma
-
\beta(a_\gamma)\gamma\otimes a_\beta
\Bigr).
\end{aligned}
\]

\noindent
The vanishing of these residues for all $i,j$ is equivalent to the condition \eqref{main identity}
on $\gamma(x)=0$, which implies the statement.

\end{proof}

The following definition introduces the key notion of an \textit{oriented $\vee$-system}. 
\begin{definition}\label{Def.Oriented V-system}
Let $\mathcal A\subset V^*$ be a finite set of non-zero, pairwise
non-collinear covectors, and let
$\Phi\colon \mathcal A\longrightarrow V\setminus\{0\}$
be a map.
For $\alpha\in\mathcal A$, set 
$a_\alpha=\Phi(\alpha).$
We say that the pair $(\mathcal A,\Phi)$ is an
\textit{oriented $\vee$-system} if for every
$\gamma\in\mathcal A$ and every two-dimensional subspace
$\pi\subset V^*$ containing $\gamma$ there exists a scalar
$\lambda_{\gamma,\pi}\in\mathbb C$ such that
\begin{equation}\label{O.A. v-system.1}
\sum_{\beta\in\pi\cap\mathcal A}
\gamma(a_\beta)\beta
=
\lambda_{\gamma,\pi}\gamma
\end{equation}
and
\begin{equation}\label{O.A. v-system.2}
\sum_{\beta\in\pi\cap\mathcal A}
\beta(a_\gamma)a_\beta
=
\lambda_{\gamma,\pi}a_\gamma.
\end{equation}
Relations \eqref{O.A. v-system.1}, \eqref{O.A. v-system.2} will be called  the
\textit{oriented $\vee$-conditions}.
\end{definition}

Let us now show that conditions \eqref{O.A. v-system.1} imply
\eqref{O.A. v-system.2} under an additional non-degeneracy assumption.
For a two-dimensional subspace $\pi\subset V^*$, set
\[
V_\pi
=
\left\langle
a_\beta\colon \beta\in\pi\cap\mathcal A
\right\rangle
\subset V.
\]

\begin{proposition}\label{Prop.pairing}
Suppose that if $|\pi\cap \mathcal A|\ge 3$ then $\dim V_\pi \le 2$ and  
the natural pairing
\begin{equation}\label{pairing}
\pi\times V_\pi\longrightarrow\mathbb C,
\qquad
(\alpha,v)\longmapsto\alpha(v),
\end{equation}
is non-degenerate if $\dim V_\pi =2$.  
Then conditions
\eqref{O.A. v-system.1} for all
$\gamma\in\pi\cap\mathcal A$ imply conditions
\eqref{O.A. v-system.2}.
\end{proposition}
\begin{proof}
If $\pi\cap\mathcal{A}=\{ \gamma \},$ then conditions \eqref{O.A. v-system.1} and \eqref{O.A. v-system.2} hold for $\lambda_{\gamma, \pi}=\gamma(a_{\gamma})$.

Suppose now that 
$\pi\cap\mathcal A=\{\beta,\gamma\}.$
%
%
The conditions \eqref{O.A. v-system.1} for $\gamma$ and $\beta$,
respectively, are
\[
\gamma(a_\gamma)\gamma+\gamma(a_\beta)\beta
=
\lambda_{\gamma,\pi}\gamma
\]
and
\[
\beta(a_\beta)\beta+\beta(a_\gamma)\gamma
=
\lambda_{\beta,\pi}\beta.
\]
Since $\beta$ and $\gamma$ are non-collinear, these identities imply
\[
\gamma(a_\beta)=0,
\qquad
\beta(a_\gamma)=0,
\]
and
\[
\lambda_{\gamma,\pi}=\gamma(a_\gamma),
\qquad
\lambda_{\beta,\pi}=\beta(a_\beta).
\]
 This implies conditions \eqref{O.A. v-system.2} for $\beta\,\, \text{and}\, \gamma .$

Now suppose that
$|\pi\cap\mathcal A|\geq3.$
 Define the linear operator
\[
T_\pi\colon \pi\longrightarrow\pi,
\qquad
T_\pi(\alpha)
=
\sum_{\beta\in\pi\cap\mathcal A}
\alpha(a_\beta)\beta,\quad \alpha\in \pi.
\]
By \eqref{O.A. v-system.1}, every
$\gamma\in\pi\cap\mathcal A$ is an eigenvector of $T_\pi$.
Since $T_\pi$ has at least three non-collinear eigenvectors in the plane $\pi$ it follows that
\[
T_\pi=\lambda_\pi\operatorname{Id}_\pi
\]
for some $\lambda_\pi\in\mathbb C$, and $\lambda_\pi =\lambda_{\gamma, \pi}\,\,\forall\gamma \in \mathcal{A}\cap \pi.$
Define now
$\widehat T_\pi\colon V_\pi\longrightarrow V_\pi$
by
\[
\widehat T_\pi(v)
=
\sum_{\beta\in\pi\cap\mathcal A}
\beta(v)a_\beta, \quad v\in V_{\pi}.
\]
For every $\alpha\in\pi$ and $v\in V_\pi$, we have
$\alpha\bigl(\widehat T_\pi(v)\bigr)=T_\pi(\alpha)(v)=\lambda_\pi \alpha(v).$

Consider firstly the case $\dim V_\pi =2$. 
It follows from the assumption that
$\widehat T_\pi
=
\lambda_\pi\operatorname{Id}_{V_\pi}$,  
which implies the required condition \eqref{O.A. v-system.2}.

Consider now the case  $\dim V_\pi =1$. Fix a basis vector $u\in V_\pi$. Then for any $\beta \in \pi$ we have $a_\beta = c_\beta u$ for some $c_\beta \in \mathbb C$. It follows that for any $\alpha \in \pi$
$$
T_\pi(\alpha) = \alpha(u) \sum_{\beta\in\pi\cap\mathcal A} c_\beta \beta. 
$$
This implies that the operator $T_\pi$ has rank at most 1, hence $\lambda_\pi=0$. It also follows that  $\alpha(u)=0$ for all $\alpha \in \pi$ or $\sum_{\beta\in\pi\cap\mathcal A} c_\beta \beta=0$. Therefore 
$$
\widehat T_\pi(u) =  \sum_{\beta\in\pi\cap\mathcal A} c_\beta \beta(u) u =0, 
$$
which implies condition \eqref{O.A. v-system.2}.
\end{proof}

We have the following corollary of Proposition \ref{Prop.pairing}. 
\begin{corollary}\label{dimension2 and OAE conditions}
    If $\dim V =2$, then conditions \eqref{O.A. v-system.1} imply conditions \eqref{O.A. v-system.2}. 
\end{corollary}

The next statement explains the importance of the oriented $\vee$-systems for the oriented associativity equations.

\begin{theorem}\label{identity=O.A.E}
 A vector potential \eqref{logarithmic solution}
satisfies the oriented associativity equations \eqref{Cp-matrix}, \eqref{O.A.E} if and only if $(\mathcal{A}, \Phi)$ is an oriented $\vee$-system.
\end{theorem}
\begin{proof}
   Firstly, we derive the condition \eqref{O.A. v-system.1} assuming that the vector potential \eqref{logarithmic solution}
satisfies the oriented associativity equations. 
By Lemma \ref{identity on L} we have 
     \begin{equation}\label{main identity with y1}
       \sum_{\beta \in \mathcal{A}\setminus \{\gamma \}} \Big( \frac{  \gamma (a_\beta) \beta (y) a_{\gamma} -  \beta (a_{\gamma}) \gamma (y)a_{\beta} }{ \beta (x)  } \Big)\gamma \wedge \beta   =0  
    \end{equation}
and 
    \begin{equation}\label{main identity with z1}
       \sum_{\beta \in \mathcal{A}\setminus \{\gamma \}} \Big( \frac{  \gamma (a_\beta) \beta (z) a_{\gamma} -   \beta (a_{\gamma}) \gamma (z) a_{\beta}}{ \beta (x)  } \Big)\gamma \wedge \beta   =0 
    \end{equation}
on the hyperplane $\Pi_{\gamma}=\{x \in V\colon \gamma(x)=0  \}$ for any $y,z \in V$.
Now, we multiply \eqref{main identity with y1} by $\gamma(z)$ and \eqref{main identity with z1} by $\gamma(y)$, and we subtract the resulting identities to get
    \begin{equation}\label{main identity with y and z 1}
       \sum_{\beta \in \mathcal{A}\setminus \{\gamma \}} { \gamma (a_\beta) \Big( \beta (y) \gamma (z) -   \gamma (y)\beta (z)\Big)} \frac{\gamma \wedge \beta }{\beta(x)}|_{\Pi_\gamma}  =0     
    \end{equation}
    since $a_{\gamma}\neq 0.$
Now, consider a two-dimensional plane $\pi \subset V^{\ast}$ which contains $\gamma$ and another vector $\beta \in \mathcal{A}$.
Let us break down the summation in relation \eqref{main identity with y and z 1} over all such two-dimensional planes $\pi$.
We have 
   \begin{equation}\label{main identity sum over planes}
     \sum_{\pi}  \sum_{\beta \in \pi \cap \mathcal{A}\setminus \{\gamma \}} { \gamma (a_\beta) \Big( \beta (y) \gamma (z) -   \gamma (y)\beta (z)\Big)} \frac{\gamma \wedge \beta }{\beta(x)}|_{\Pi_\gamma}  =0.     
    \end{equation}
It is easy to see that $\frac{\gamma \wedge \beta }{\beta(x)}|_{\Pi_\gamma} $ is independent of the choice of the vector  $\beta \in \pi\cap \mathcal{A} \setminus \{\gamma\}$.
Note also that all the sums for different two-dimensional planes $\pi$ have singularity on different hyperplanes in $\Pi_{\gamma}.$
It follows that we get the following equivalent form of identity \eqref{main identity sum over planes}:
\begin{equation}\label{simplified identity}
  \sum_{\beta \in \pi \cap \mathcal{A}} { \gamma (a_\beta) \Big( \beta (y) \gamma (z) -   \gamma (y)\beta (z)\Big)} =0.   
\end{equation}
Relation \eqref{simplified identity} can be written equivalently as
$$\sum_{\beta \in \pi \cap \mathcal{A}} { \gamma (a_\beta) \beta  \wedge \gamma  } =0,$$
 which shows that condition \eqref{O.A. v-system.1} is satisfied.
 
 Now let us prove relations \eqref{O.A. v-system.2}. 
 Consider again a two-dimensional plane $\pi$ containing fixed $\gamma \in \mathcal{A}$ together with an additional covector $\beta \in \mathcal A$. 
 Then splitting the sum in relation \eqref{main identity} over all such planes gives
\begin{equation}\label{main identity sum over planes 2}
     \sum_{\pi} \sum_{\beta \in \pi \cap \mathcal{A}} \Big({ \gamma (a_\beta)  \beta \otimes a_\gamma -\beta(a_{\gamma})   \gamma \otimes a_\beta \Big)}\frac{\gamma \wedge \beta }{\beta(x)}|_{\Pi_{\gamma}}  =0.    
    \end{equation}
Similarly to above, relation \eqref{main identity sum over planes 2} implies that
\begin{equation}
    \sum_{\beta \in \pi \cap \mathcal{A}} { \gamma (a_\beta)  \beta \otimes a_\gamma 
    =\sum_{\beta \in \pi \cap \mathcal{A}}\beta(a_{\gamma})   \gamma \otimes a_\beta}. 
\end{equation}
Since we established that $\sum_{\beta \in \pi \cap \mathcal{A}}\gamma (a_{\beta})\beta =\lambda \gamma$
for some $\lambda \in \mathbb{C}$,
we get that
\begin{equation}
   \sum_{\beta \in \pi \cap \mathcal{A}}\beta(a_{\gamma})   \gamma \otimes a_\beta = \lambda \gamma\otimes a_\gamma,
\end{equation}
which shows that condition \eqref{O.A. v-system.2} is satisfied, and $(\mathcal{A}, \Phi)$ is an oriented $\vee$-system.

Assume now that $(\mathcal{A}, \Phi)$ is an oriented $\vee$-system. Let us fix a covector $\gamma \in \mathcal{A}$ and consider the corresponding hyperplane $\Pi_{\gamma}$. Let
  \begin{equation}\label{main identity.1}
        S_{\gamma}= \sum_{\beta \in \mathcal{A}} \Big( \frac{ \gamma (a_\beta) \beta \otimes a_{\gamma} - \beta (a_{\gamma}) \gamma \otimes a_{\beta}  }{ \beta (x)  } \Big)\gamma \wedge \beta.      
    \end{equation}
Consider a two-dimensional plane $\pi$ containing $\gamma$ and another covector from $\mathcal A$. Then splitting the sum in $S_{\gamma}$ over all such planes gives
\begin{equation}\label{main identity sum over planes 3}
   S_{\gamma}=   \sum_{\pi} \sum_{\beta \in \pi \cap \mathcal{A}} \frac{\gamma \wedge \beta }{\beta(x)} \Big({ \gamma (a_\beta)  \beta \otimes a_\gamma -\beta(a_{\gamma})   \gamma \otimes a_\beta \Big)}.
     \end{equation}
Since $\frac{\gamma \wedge \beta }{\beta(x)} |_{\Pi_{\gamma}}$ does not depend on $\beta\in \pi\cap \mathcal{A},$ it can be taken out of the internal sum in \eqref{main identity sum over planes 3}. 
The oriented $\vee$-conditions \eqref{O.A. v-system.1}
and \eqref{O.A. v-system.2} then imply that the internal sum and hence $S_{\gamma}$ vanish on $\Pi_{\gamma}.$
Lemma \ref{identity on L} implies that the oriented associativity equations   for the vector potential \eqref{logarithmic solution} hold,
which completes the proof.
\end{proof}
\section{The planar case}\label{section.Examples on the plane}

In this section we list all the oriented $\vee$-systems on the plane $V\cong \mathbb{C}^2$.
We will also determine in examples when such solutions arise from a single scalar prepotential.

Any configuration with one (non-zero) covector in $V^*\cong \mathbb{C}^2$
and one (non-zero) vector in $V\cong \mathbb{C}^2$ is an oriented $\vee$-system.

In the case of a two-element set $\mathcal{A}$ we can assume that the covectors are $e^1, e^2$.
The corresponding vectors $a_1=(a_1^1,a_1^2),\, a_2=(a_2^1, a_2^2) \in \mathbb C^2$ produce an oriented $\vee$-system $\{(e^1,a_1),(e^2, a_2) \}$ if and only if $a_1^2=a_2^1=0.$
\begin{proposition}\label{Prop.general2D.vconditions}
Consider a finite configuration of
pairwise non-collinear nonzero covectors
\[
\mathcal{A}=\{\alpha_1,\ldots,\alpha_k\}\subset V^*,\quad k \geq 3,
\]
where
\[
\alpha_i=\alpha_i^1 e^1+\alpha_i^2 e^2,
\qquad i=1,\ldots,k.
\]
Associate to each \(\alpha_i\) a nonzero vector
\[
a_i=a_i^1e_1+a_i^2e_2 \in V.
\]
Then \((\mathcal{A},\Phi)\), where \(\Phi(\alpha_i)=a_i\), is an
oriented \(\vee\)-system if and only if
\begin{equation}\label{general-2D-system}
\begin{aligned}
\sum_{i=1}^{k}\alpha_i^2 a_i^1&=0,\\
\sum_{i=1}^{k}\alpha_i^1 a_i^2&=0,\\
\sum_{i=1}^{k}\alpha_i^1 a_i^1
&=
\sum_{i=1}^{k}\alpha_i^2 a_i^2.
\end{aligned}
\end{equation}
\end{proposition}

\begin{proof}
By Corollary \ref{dimension2 and OAE conditions}, the oriented
\(\vee\)-conditions are equivalent to
\begin{equation}\label{general-2D-eigencondition}
\sum_{\beta\in\mathcal A}\gamma(a_\beta)\beta
=
\lambda\gamma,
\qquad
\gamma\in\mathcal A,
\end{equation}
for some \(\lambda\in\mathbb C\) which is independent of
\(\gamma\).

Since the left-hand side of
\eqref{general-2D-eigencondition} depends linearly on \(\gamma\),
and since \(\mathcal A\) spans \(V^*\), condition
\eqref{general-2D-eigencondition} is equivalent to requiring it
for a basis of \(V^*\).

For \(\gamma=e^j,\,\, j=1,2\), 
the condition \eqref{general-2D-eigencondition} takes the form
\[
\sum_{i=1}^{k}
a_i^j(\alpha_i^1e^1+\alpha_i^2e^2)
=
\lambda e^j.
\]
By considering both components of each of the two equations and eliminating $\lambda$ we get the statement.
\end{proof}
Let us now consider an example where the configuration of covectors is
\begin{equation}\label{B2-configuration}
    \mathcal{A}=B_2^+=\{e^1,e^2,e^1+e^2,e^1-e^2\},
\end{equation}
By Proposition \ref{Prop.general2D.vconditions} we can parametrize the corresponding nonzero vectors as

\begin{align}\label{B2-parametrization}
   a_1&=(t_1,t_2),
\qquad
a_2=(t_4,t_5), \nonumber\\
a_3&=
\left(
t_3,
\frac{t_1-t_2+2t_3 +t_4-t_5}{2}
\right),\\ 
a_4&=
\left(
t_3+t_4,
\frac{-t_1-t_2-2t_3-t_4+t_5}{2}
\right),\nonumber
\end{align}
where $t_i \in \mathbb{C},\, (1\leq i\leq 5)$ are such that all vectors \eqref{B2-parametrization} are non-zero.

Let us clarify which configurations of this type have the property that there exists a single prepotential defining the vector potential. 
More specifically, consider the vector potential 
\begin{equation}\label{solution ci}
\begin{aligned}
f^i=
a_1^i\,x_1\log x_1
+a_2^i\,x_2\log x_2
+a_3^i\,(x_1+x_2)\log(x_1+x_2) 
+
a_4^i\,(x_1-x_2)\log(x_1-x_2),
\end{aligned}
\end{equation}
where $i=1,2$
and let us define the prepotential
\begin{equation}\label{prep F}
\begin{aligned}
\mathcal F=
A\,x_1^2\log x_1
+B\,x_2^2\log x_2
+C\,(x_1+x_2)^2\log(x_1+x_2)+
D\,(x_1-x_2)^2\log(x_1-x_2),
\end{aligned}
\end{equation}
where $A, B, C, D \in \mathbb{C}.$

Let $\eta=(\eta_{ij})_{i,j=1}^{2}$
be defined as 
\begin{equation*}
    \eta_{ij}=\frac{1}{2}\sum_{k=1}^2 x_k \frac{\partial^3 \mathcal F}{\partial x_i \partial x_j \partial x_k}=\sum_{\alpha\in\widetilde B_2}
\alpha_i\alpha_j ,
\end{equation*}
where 
$$\widetilde B_2=
\{\sqrt A\,e^1,\sqrt B\,e^2,
\sqrt C(e^1+e^2),
\sqrt D(e^1-e^2)\}.$$
Then the inverse matrix $\eta^{-1}= (\eta^{ij})^2_{i,j=1}$
has the form
\begin{equation}\label{metric for B2}
\eta^{-1}
=
\frac1{\Delta}
\begin{pmatrix}
B+C+D & D-C\\
D-C & A+C+D
\end{pmatrix},    
\end{equation}
where
\begin{equation}\label{non-degeneracy}
    \Delta
=
AB+(A+B)(C+D)+4CD
\end{equation}
is assumed to be non-zero.
Let us investigate whether there exists a function \eqref{prep F} such that 

\begin{equation}\label{Ci from F}
f^i=\eta^{i1}
\frac{\partial\mathcal F}{\partial x_1}
+ 
\eta^{i2}
\frac{\partial\mathcal F}{\partial x_2}
+
\ell^i(x),
\end{equation}
where $\ell^i$ is a linear function.

\begin{proposition}\label{Prop.B2.Existence of F}
Consider the five-parameter family \eqref{B2-parametrization} of oriented $\vee$-systems with $\mathcal{A}=B_2^+$.
Assume that
\[
t_2t_4(t_2-t_4)(t_1t_4-t_2t_5)(t_1t_5-t_2t_4)\neq0.
\]
Then the logarithmic vector potential \eqref{solution ci} arises from
a single prepotential of the form \eqref{prep F} through
\eqref{Ci from F} if and only if the following two compatibility
conditions hold:
\begin{equation}\label{condition 1 for B2}
\det
\begin{pmatrix}
t_3 & t_1+t_2 & 0\\
-(t_1-t_2+t_4-t_5) & 2t_1 & 2t_2\\
0 & t_5 & t_4
\end{pmatrix}
=0,
\end{equation}
and
\begin{equation}\label{condition 2 for B2}
    (t_2-t_4)(t_1t_5-t_2t_4)=2(t_2t_5-t_1t_4).
\end{equation}
In this case, \(A\neq0\), and once \(A\) is fixed, the remaining
coefficients of the prepotential are uniquely determined by
\begin{align}
B&=\frac{t_4}{t_2}A,
\label{B2.case1.B.new}\\[2mm]
C&=
\frac{
At_4(t_1-t_2+t_4-t_5)
}{
2(t_2t_5-t_1t_4)
},
\label{B2.case1.C.new}\\[2mm]
D&=
\frac{
At_4(t_1+t_2-t_4-t_5)
}{
2(t_2t_5-t_1t_4)
}.
\label{B2.case1.D.new}
\end{align}
The determinant of the corresponding metric $\eta$ is
\begin{equation}\label{delta}
\Delta
=
A^2
\frac{
t_4(t_2-t_4)^2(t_1t_5-t_2t_4)
}{
t_2(t_2t_5-t_1t_4)^2
}.    
\end{equation}
\end{proposition}
\begin{proof}
Set $\mu=\frac{2}{\Delta}.$
Equations \eqref{solution ci}, \eqref{prep F} and \eqref{Ci from F} are equivalent to the following conditions:

\begin{align}
a_1^1&=\mu A(B+C+D),\label{a11}\\
a_2^1&=\mu B(D-C),\label{a21}\\
a_3^1&=\mu C(B+2D),\label{a31}\\
a_4^1&=\mu D(B+2C),\label{a41}\\
a_1^2&=\mu A(D-C),\label{a12}\\
a_2^2&=\mu B(A+C+D),\label{a22}\\
a_3^2&=\mu C(A+2D),\label{a32}\\
a_4^2&=-\mu D(A+2C).\label{a42}
\end{align}

Suppose that these equations hold.
Since
$t_2=a_1^2=\mu A(D-C)\neq 0$ it follows that
$A\neq 0.$
We first solve a subsystem of
\eqref{a11}--\eqref{a42}.
From \eqref{a21} and \eqref{a12}, we get \eqref{B2.case1.B.new}.

From equations \eqref{a11} and \eqref{a22} we get
\begin{equation}\label{mu}
    \mu A^2=
\frac{t_2(t_2t_5-t_1t_4)}
{t_4(t_2-t_4)}.
\end{equation}
Substituting this into equations \eqref{a21} and \eqref{a11} we obtain \eqref{B2.case1.C.new} and \eqref{B2.case1.D.new}.

Let us now substitute the found expressions for $\mu, B, C \,\text{and} \, D$ into \eqref{a31} we get
\[
\mu C(B+2D)-t_3
=
-\frac{
P(t_1,t_2,t_3,t_4,t_5)
}{
2(t_1t_4-t_2t_5)
}=0,
\]
where
\[
\begin{aligned}
P(t_1,t_2,t_3,t_4,t_5)
= t_4 (t_1 +t_2)(t_1 -t_2 +t_4-t_5)-2t_3(t_2 t_5-t_1 t_4).
\end{aligned}
\]
Since
\[
\begin{aligned}
P(t_1,t_2,t_3,t_4,t_5)
={}&
\det
\begin{pmatrix}
t_3 & t_1+t_2 & 0\\
-(t_1-t_2+t_4-t_5) & 2t_1 & 2t_2\\
0 & t_5 & t_4
\end{pmatrix}
\end{aligned}
\]
we get condition
\eqref{condition 1 for B2}.
Substituting \eqref{B2.case1.B.new}--\eqref{B2.case1.D.new} into \eqref{non-degeneracy} gives \eqref{delta}. 
By combining \eqref{non-degeneracy}, \eqref{mu} and $\mu=\frac{2}{\Delta}$, and canceling the nonzero common factors, we get \eqref{condition 2 for B2}.

Conversely, one can check that relations \eqref{condition 1 for B2}--\eqref{B2.case1.D.new} imply formulas \eqref{a11}--\eqref{a42}.
This completes the proof.
\end{proof}
\begin{remark}
One can weaken the requirement of existence of a single prepotential by allowing that equality \eqref{Ci from F} holds up to a constant proportionality factor. 
In this case \eqref{condition 1 for B2} becomes the only required compatibility condition.
\end{remark}
Similar analysis can be done for the configuration 
$A_2^+=\{e^1 , e^2, e^1+e^2\}$.
The parameterization of the corresponding vectors that define an oriented $\vee$-system can be obtained from the parameterization \eqref{B2-parametrization} by imposing $a_4=0$. 
It takes the form 
\[
a_1=(t_1,t_2),\quad a_2=(t_3,t_1+t_2-t_3),\qquad a_3=(-t_3,-t_2).\]
This oriented $\vee$-system admits a single prepotential if and only if 
$t_1=t_3 +2$ 
under the assumption that $t_2 t_3(t_1 +t_2)\neq 0.$
The compatibility condition $t_1=t_3 +2$ is replaced with $t_1 \ne t_3$ if we allow a coefficient of proportionality in the condition \eqref{Ci from F}.

\section{Subsystems}
In this section, we introduce subsystems of oriented $\vee$-systems
and show that every such subsystem is itself an oriented
$\vee$-system. By Theorem~\ref{identity=O.A.E}, this construction
provides new logarithmic solutions of the oriented associativity
equations from a given oriented $\vee$-system.

Let $(\mathcal A,\Phi)$ be an oriented $\vee$-system, where
\[
\mathcal A\subset V^*,
\qquad
\Phi:\mathcal A\longrightarrow V\setminus\{0\}.
\]
Let $\mathcal B\subset\mathcal A$ and assume that
$\mathcal B=\mathcal A\cap W$ for a linear subspace 
$W=\langle \beta \colon \beta \in\mathcal B\rangle\subset V^*.$
In this case we say that $\mathcal B$ is {\it a subsystem} of $\mathcal A$. 

For every $\beta\in\mathcal B$, we retain the vector associated with
$\beta$ in the original system, that is,
\[
\Phi_{\mathcal B}(\beta)
:=
\Phi(\beta)=a_\beta.
\]
Equivalently,
$\Phi_{\mathcal B}=\Phi|_{\mathcal B}.$

The following statement takes place.
\begin{theorem}\label{subsystem solution}
The pair $(\mathcal{B}, \Phi|_{\mathcal{B}})$
is an oriented $\vee$-system.
\end{theorem}

\begin{proof}
Since $(\mathcal{A},\Phi)$ is an oriented $\vee$-system, for every
$\gamma\in\mathcal{A}$ and every two-dimensional plane
$\pi\subset V^*$ containing $\gamma$, there exists
$\lambda_{\gamma,\pi}\in\mathbb{C}$ such that
\begin{equation}\label{O.A.E on A.1.a}
\sum_{\beta\in\pi\cap\mathcal{A}}
\gamma(a_\beta)\,\beta
=
\lambda_{\gamma,\pi}\gamma
\end{equation}
and
\begin{equation}\label{O.A.E on A.2.b}
\sum_{\beta\in\pi\cap\mathcal{A}}
\beta(a_\gamma)\,a_\beta
=
\lambda_{\gamma,\pi}a_\gamma.
\end{equation}


Let $\gamma\in\mathcal{B}$, and consider a two-dimensional plane
$\widetilde{\pi}\subset W$ containing $\gamma$. 
Since $\mathcal{B}=\mathcal{A}\cap W$
and $\widetilde{\pi}\subset W$, we have
$ \widetilde{\pi}\cap\mathcal{B}
=
\widetilde{\pi}\cap\mathcal{A}.$
Therefore, applying \eqref{O.A.E on A.1.a} to the plane
$\pi=\widetilde{\pi}$ gives
\begin{equation}\label{O.A.E on B.1}
\begin{aligned}
\sum_{\beta\in\widetilde{\pi}\cap\mathcal{B}}
\gamma(a_\beta)\,\beta
&=
\sum_{\beta\in\widetilde{\pi}\cap\mathcal{A}}
\gamma(a_\beta)\,\beta
=
\lambda_{\gamma,\widetilde{\pi}}\,\gamma.
\end{aligned}
\end{equation}
Similarly, by \eqref{O.A.E on A.2.b},
\begin{equation}\label{O.A.E on B.2}
\begin{aligned}
\sum_{\beta\in\widetilde{\pi}\cap\mathcal{B}}
\beta(a_\gamma)\,a_\beta
&=
\sum_{\beta\in\widetilde{\pi}\cap\mathcal{A}}
\beta(a_\gamma)\,a_\beta
=
\lambda_{\gamma,\widetilde{\pi}}a_\gamma.
\end{aligned}
\end{equation}
Relations \eqref{O.A.E on B.1} and \eqref{O.A.E on B.2} are precisely
the oriented $\vee$-conditions for
$(\mathcal{B},\Phi|_{\mathcal{B}})$. 
\end{proof}
We now introduce a reduced version of the subsystem construction,
which takes into account that
$W=\langle\mathcal B\rangle\subset V^*$
might have dimension smaller than the dimension of $V$.

Let
$L=\bigcap_{\beta\in\mathcal B}\Pi_\beta\subset V,$
and choose a complementary subspace $M\subset V$ such that
\[
V=M\oplus L.
\]
Then every vector $v\in V$ admits a unique decomposition
\begin{equation}\label{direct sum for a}
v=v^M+v^L,
\qquad
v^M\in M,\quad v^L\in L.
\end{equation}
Note that for any $\beta \in \mathcal{B}$
\[
\beta(v^L)=0
\qquad
\text{for all } v\in V.
\]
In particular, 
$\beta(a_\beta^L)=0$
and hence
$\beta(a_\beta)=\beta(a_\beta^M).$

For the subsystem
$\mathcal B=\mathcal A\cap W,$
consider the restrictions
\[
\mathcal B|_M
=
\{\beta|_M:\beta\in\mathcal B\}
\subset M^*.
\]
These restrictions are pairwise distinct, that is different elements of $\mathcal{B}$ restrict to different elements in $\mathcal{B}|_{M}$.
Indeed, if
\[
\beta|_M=\gamma|_M
\]
for some $\beta,\gamma\in\mathcal B$, then it follows that $\beta=\gamma.$

We can therefore define a map
\[
\Psi\colon\mathcal B|_M\longrightarrow M
\]
by
\[
\Psi(\beta|_M)
=
a_\beta^M
=
(\Phi(\beta))^M,
\qquad
\beta\in\mathcal B.
\]

The following theorem allows to obtain oriented $\vee$-systems from a given one. 
\begin{theorem}
The pair $(\mathcal{B}|_{M}, \Psi)$
is equivalent to an oriented $\vee$-system.
\end{theorem}

\begin{proof}
The oriented $\vee$-system conditions for $(\mathcal{A},\Phi)$ are
\begin{equation}\label{O.A.E on A.1.a.1}
 \sum_{\beta \in \pi \cap \mathcal{A}} { \gamma (a_\beta) \beta =\lambda_{\gamma,\pi} \gamma} 
\end{equation}
and 
\begin{equation}\label{O.A.E on A.2.b.1}
       \sum_{\beta \in \pi \cap \mathcal{A}} {\beta (a_\gamma) a_\beta =\lambda_{\gamma,\pi} a_\gamma} 
  \end{equation} 
for any two-dimensional plane $\pi \subset V^{\ast}, \, \gamma\in \pi\cap \mathcal{A} $
and some $\lambda_{\gamma,\pi} \in \mathbb{C}$.
\medskip

Let 
us fix $\gamma \in \mathcal{B}$.
Consider any two-dimensional plane $\widetilde{\pi}=\langle \gamma,\delta\rangle \subset W$ with $\delta \in \mathcal{B}$, $\delta \neq \gamma$.
Then
\[
\widetilde{\pi} \cap \mathcal{A} = \widetilde{\pi} \cap \mathcal{B},
\]
and hence
\begin{equation}\label{restricted identity corrected.1}
\sum_{\delta\in\widetilde{\pi}\cap\mathcal{B}}
\gamma(a_{\delta})\,\delta=\lambda_{\gamma,\widetilde{\pi}}\gamma.
\end{equation}
Also we have
\begin{equation}\label{restricted identity corrected.2}
       \sum_{\delta \in \widetilde{\pi} \cap \mathcal{B}} {\delta(a_\gamma) a_\delta =\lambda_{\gamma,\widetilde{\pi}} a_\gamma}. 
  \end{equation} 
We now use the decomposition $a_\delta = a_\delta^L + a_\delta^M$
and the property
\begin{equation}\label{decomposition relation}
  \gamma(a_\delta)=\gamma(a_\delta^M).  
\end{equation}

By substituting \eqref{decomposition relation} into \eqref{restricted identity corrected.1}, we obtain
\begin{equation}\label{rest 1}
    \sum_{\delta\in\widetilde{\pi}\cap\mathcal{B}}
\gamma(a_{\delta}^M)\,\delta|_M=\lambda_{\gamma,\widetilde{\pi}} \gamma|_M.
\end{equation}
 Also we have
\begin{equation}\label{rest 2}
   \sum_{\delta \in \widetilde{\pi} \cap \mathcal{B}} {\delta(a_\gamma^M) a_\delta^M =\lambda_{\gamma,\widetilde{\pi}} a_\gamma^M}. 
\end{equation}

\medskip

Note that noncollinear covectors $\delta \in \mathcal B$ correspond to noncollinear covectors $\delta|_M$, and that if a vector $a_\delta^M =0$ then covector $\delta$ has zero contribution to the sums in \eqref{rest 1}, \eqref{rest 2} hence can be omitted in the considerations. 

Consider a two-dimensional plane $\widehat{\pi}\subset M^*$ spanned by a pair of covectors $\beta_1 |_M , \beta_2|_M,$ where $\beta_1 , \beta_2 \in \mathcal{B}.$
Let $\widetilde{\pi}=\{\alpha\in W\colon \alpha|_M \in \widehat{\pi} \}. $
Note that $\dim \widetilde{\pi}=2$ since if an element $\alpha_0 \in \widetilde{\pi}$ has a zero restriction to $M$ then it must be that $\alpha_0 =0$.
Since the restriction map $\delta \to \delta|_M$ is an isomorphism  between $\widetilde{\pi}$ and $\widehat{\pi}$ the sums in \eqref{rest 1} and \eqref{rest 2} can be written as sums over elements in $\widehat{\pi}\cap \mathcal{B}|_M$ which gives the oriented $\vee$-conditions for $(\mathcal{B}|_M , \Psi)$.
This completes the proof.
\end{proof}


 \section{Restrictions}\label{Restriction of solutions}

In this section we show how to obtain new solutions of the oriented associativity equations from known oriented $\vee$-systems 
by considering their restrictions. 

Let $(\mathcal{A}, \Phi)$ be an oriented $\vee$-system, and let $\mathcal B \subset \mathcal A$ be a subsystem.
For \(\alpha\in\mathcal A\), put
$\Pi_\alpha=\{x\in V:\alpha(x)=0\},$
and define
\[
L=L_{\mathcal B}
=
\bigcap_{\beta\in\mathcal B}\Pi_\beta.
\]
Then \(\operatorname{codim}L=m\), where $m = \dim W$, $W= \langle \beta: \beta \in \mathcal B\rangle$. 

For each \(\alpha\in\mathcal A\setminus\mathcal B\), the restriction
$\alpha|_L\in L^*$
defines a hyperplane
\[
\Pi_\alpha\cap L
=
\{x\in L:\alpha(x)=0\}
\]
in \(L\). 
We denote by
\[L_\Sigma
=
L\setminus
\bigcup_{\alpha\in\mathcal A\setminus\mathcal B}
(\Pi_\alpha\cap L)
=
\{x\in L:\alpha(x)\neq0
\text{ for all }\alpha\in\mathcal A\setminus\mathcal B\}
\]
the complement of these hyperplanes in \(L\).
Let also 
\[
V_\Sigma = V \setminus 
\bigcup_{\alpha\in\mathcal A}\Pi_\alpha.
\]

Choose a complement \(M\) of \(L\), so that
\[V=L\oplus M.\]
For every \(a_\alpha\in V\) there is a unique
decomposition
\[
a_\alpha=a_\alpha^L+a_\alpha^M,
\qquad
a_\alpha^L\in L,\quad a_\alpha^M\in M.
\]


Recall that it defines an associative multiplication 

\begin{equation}\label{product2}
    u\ast v
    =
    \sum_{\alpha\in\mathcal A}
    \frac{\alpha(u)\alpha(v)}{\alpha(x)}\,a_\alpha,
\end{equation}
where $a_\alpha =\Phi (\alpha), \, u,v\in T_x V,$ and $ x\in V_\Sigma.$

Next, we study the behavior of this multiplication as \(x\) approaches
a point \(x_0\in L_\Sigma\). In particular, we shall show that for
tangent vectors \(u_0,v_0\in T_{x_0}L\), the singular contributions
corresponding to the covectors in \(\mathcal B\) disappear and that the resulting multiplication is well-defined
and closed on \(T_{x_0}L\).

 
 Let us extend vectors $u_{0}$ and $v_{0}$ to two local analytic vector fields $u(x), v(x)$ in the neighbourhood $U$ of $x_{0}$ such that they are tangential to $\Pi_\beta$ on $\Pi_\beta$ for each $\beta \in \mathcal B$, and $u_{0}=u(x_{0})$, $v_{0}=v(x_{0})$. 
%
We want to study the limit of $u(x)\ast v(x)$ when $x$ tends to $x_0 \in L$. 
\begin{lemma}\label{the limit of * exists}
The limit of the product $u(x) \ast v(x)$ exists when vector $x$ tends to $x_{0}\in L $ and it satisfies
\begin{equation}\label{limit formula}
u_{0} \ast v_{0} =\sum_{\alpha \in \mathcal{A}\setminus \mathcal{B}} \frac{\alpha(u_0)\alpha(v_0)}{\alpha(x_0)}a_{\alpha}. 
\end{equation}
In particular, the product $u_{0} \ast v_{0}$ is determined by vectors $u_{0}$ and $v_{0}$ only. 
\end{lemma}
We omit the proof as it is fully analogous to the proof of the corresponding statement for the usual $\vee$-systems  (cf. \cite{MishaVeselov2007}).

\begin{lemma}\label{closed multiplication}
For any $x\in L_\Sigma$ and
$u,v\in T_xL\cong L,$
the product $u\ast v \in T_xL$. Thus,
\[
\ast:T_xL\times T_xL\longrightarrow T_xL.
\]
\end{lemma}

\begin{proof}
Let $\gamma\in \mathcal{B}.$
It is enough to prove that
$\gamma(u\ast v)=0.$

We have
\[
\gamma(x)=\gamma(u)=\gamma(v)=0.
\]
By Lemma~\ref{identity on L}, for any $y\in V$,
\begin{equation}\label{main identity with y}
\sum_{\beta\in\mathcal{A}\setminus\{\gamma\}}
\left(
\frac{
\gamma(a_\beta)\beta(y) a_\gamma
-
\beta(a_\gamma)\gamma(y) a_\beta
}
{\beta(x)}
\right)
\gamma\wedge\beta|_{\Pi_\gamma}
=0.
\end{equation}
Similarly, for any $z\in V$,
\begin{equation}\label{main identity with z}
\sum_{\beta\in\mathcal{A}\setminus\{\gamma\}}
\left(
\frac{
\gamma(a_\beta)\beta(z) a_\gamma
-
\beta(a_\gamma)\gamma(z) a_\beta
}
{\beta(x)}
\right)
\gamma\wedge\beta|_{\Pi_\gamma}
=0.
\end{equation}

Multiplying \eqref{main identity with y} by $\gamma(z)$ and
\eqref{main identity with z} by $\gamma(y)$, and then subtracting,
we obtain
\begin{equation}\label{main identity with y and z}
\sum_{\beta\in\mathcal{A}\setminus\{\gamma\}}
\frac{
\gamma(a_\beta)
\bigl(
\beta(y)\gamma(z)-\gamma(y)\beta(z)
\bigr)a_\gamma
}
{\beta(x)}
\gamma\wedge\beta|_{\Pi_\gamma}
=0.
\end{equation}

Evaluating the two-form $\gamma\wedge\beta$ on vectors
$a,b\in V$ gives
\begin{equation}\label{main identity with abyz}
\sum_{\beta\in\mathcal{A}\setminus\{\gamma\}}
\frac{
\gamma(a_\beta)
\bigl(
\beta(y)\gamma(z)-\gamma(y)\beta(z)
\bigr)
\bigl(
\gamma(a)\beta(b)-\gamma(b)\beta(a)
\bigr)a_\gamma
}
{\beta(x)}|_{\Pi_\gamma}
=0.
\end{equation}

Now take
$b=u$ and $y=v.$
Since $u,v\in L\subset\ker\gamma$, we have 
$\gamma(b)=\gamma(y)=0.$
Hence \eqref{main identity with abyz} reduces to
\[
\gamma(a)\gamma(z) a_\gamma
\sum_{\beta\in\mathcal{A}\setminus\{\gamma\}}
\frac{
\gamma(a_\beta)\beta(u)\beta(v)
}
{\beta(x)}|_{\Pi_{\gamma}}
=0.
\]

Choose $a,z\in V$ such that
$\gamma(a)\neq0$ and $\gamma(z)\neq0.$
Since $a_\gamma\neq0$, it follows that
\begin{equation}\label{gamma product zero}
\sum_{\beta\in\mathcal{A}\setminus \mathcal{B}}
\frac{
\gamma(a_\beta)\beta(u)\beta(v)
}
{\beta(x)}
=0
\end{equation}
as needed.
\end{proof}

\begin{theorem}\label{restricted v-system}
Let $(\mathcal A, \Phi)$ be an oriented $\vee$-system with a subsystem  $\mathcal B \subset \mathcal A$. 
Consider a collection of covectors $\mathcal{A}_L \subset L^*$ defined as restrictions $\alpha|_L , \alpha \in \mathcal{A}\setminus \mathcal{B}.$
    Define 
    $\Phi_L\colon \mathcal{A}_L \longrightarrow L$ as
    $$\Phi_L (\gamma)= \sum_{\substack{\alpha \in \mathcal{A}\\ \alpha|_{L} =\gamma}} a_{\alpha}^L.$$
    Then $(\mathcal{A}_L ,\Phi_L)$ is equivalent to an oriented $\vee$-system in $L$.
\end{theorem}

\begin{proof}
   By Lemma~\ref{the limit of * exists} the product \eqref{limit formula} is associative. By Lemma~\ref{closed multiplication} it can be rewritten as
   \begin{equation}\label{product with projection}
u\ast v
=
\sum_{\alpha\in\mathcal A\setminus\mathcal B}
\frac{\alpha(u)\alpha(v)}
     {\alpha(x)} a_\alpha
=
\sum_{\alpha\in\mathcal A\setminus\mathcal B}
\frac{\alpha(u)\alpha(v)}
     {\alpha(x)}\,a_\alpha^L
     =
     \sum_{\gamma \in \mathcal{A}_L}\frac{\gamma(u)\gamma(v)}
     {\gamma(x)}\Phi_L(\gamma),
 \end{equation}
 where $u, v \in T_x L,\, x\in L_\Sigma.$
 As we replace a configuration $(\mathcal{A}_L ,\Phi_L)$ by an equivalent configuration, the corresponding expression \eqref{product with projection} does not change.
 The statement follows by Proposition \ref{asossiative=oriented} and Theorem \ref{identity=O.A.E}.
\end{proof}
Note that different choices of $M$ give equivalent oriented $\vee$-systems. 
This follows from the fact that the corresponding associative multiplication does not depend on $M$, hence the second order derivatives of the corresponding vector potentials also do not depend on $M$.
Therefore, vector potentials differ by linear vector-valued functions which implies that the corresponding oriented $\vee$-systems are equivalent.
Furthermore, this also implies that the configuration $(\mathcal{A}_L,\widetilde{\Phi})$, where $\widetilde{\Phi}(\gamma)=\sum_{\substack{\alpha \in \mathcal{A}\\ \alpha|_{L} =\gamma}} a_{\alpha}$, 
can be transformed to the same oriented $\vee$-system in $L$ by elementary operations if we extend the elementary operations to the bigger space $V\supset L.$


Let us also give a formula for vector potential $f\colon L_\Sigma \to L$ associated with the above restriction procedure and which does not depend on the complement subspace $M$. 

We first introduce some notations. Let $\gamma_1, \dots, \gamma_m $ be a basis in $W$. 
Let $\pi_{m+1}$ be an $(m+1)$-dimensional space containing $W$ such that $\pi_{m+1}=\langle \gamma_1, ..., \gamma_m, \beta_0\rangle$, where $\beta_0 \in \mathcal{A}\setminus \mathcal{B}$. For any $\beta \in \pi_{m+1}$
let $\beta =\sum_{i=1}^m d_i \gamma_i + d_{\beta} \beta_0$ for some $d_i, d_{\beta}\in \mathbb{C}.$
Then for $x \in L_\Sigma$ we have $\beta(x) = d_\beta \beta_0(x)$, and we choose branches of logarithms such that $\log \beta(x) = \log \beta_0(x) + \log d_\beta$. 

\begin{proposition}
\label{restriction_formula_h}
Define function
\begin{equation}
f(x)=\sum_{\beta \in \mathcal{A}\setminus \mathcal{B}} \beta(x)\log \beta(x) a_{\beta} -h,\qquad x\in L_\Sigma,  
\end{equation}
where
$$h= \sum_{\pi_{m+1}} \sum_{\beta\in (\mathcal{A} \setminus \mathcal{B})\cap\pi_{m+1}}  \beta (x) \log d_{\beta} \,  a_{\beta},  $$
and summation runs over all $(m+1)$-dimensional spaces which contain $W$ and a vector from $\mathcal{A}\setminus\mathcal{B}$.
Then $f$ takes values in $L$ and satisfies the oriented associativity equations.
\end{proposition}
\begin{proof}
Since $h$ is linear it follows from Theorem \ref{restricted v-system} that $f$ satisfies the oriented associativity equations.
We have to show that $\gamma (f)=0$ for all $\gamma \in \mathcal{B}.$

Observe that $\beta(x) = d_\beta \beta_0(x)$ for any $x\in L$. Therefore it is sufficient to establish that
\begin{equation}
    \label{pim}
\sum_{\beta \in (\mathcal{A}\setminus\mathcal{B})\cap \pi_{m+1}} \gamma(a_\beta) \beta(x) =0
\end{equation}
for any $x\in L$. 
Note that the left-hand side of \eqref{pim} can be rearranged as
\begin{equation}
    \label{pim2}
\sum_{\beta \in \mathcal{A}\cap \pi_{m+1}} \gamma(a_\beta) \beta(x) =
\sum_{\pi}
\sum_{\beta \in \mathcal{A}\cap \pi} \gamma(a_\beta) \beta(x),
\end{equation}
where $\pi$ is a two-dimensional subspace of $\pi_{m+1}$ which contains $\gamma$ and another element of $\mathcal A$. For every $\pi$ the internal sum in \eqref{pim2} vanishes by the condition \eqref{O.A. v-system.1}, which implies \eqref{pim}.
\end{proof}

Let us explain that the function $f$ modulo linear terms does not depend on the choice of a basis vector $\beta_0$ in the space $\pi_{m+1}$. Indeed, 
consider an element $\widetilde{\beta}_0 \in (\mathcal{A}\setminus\mathcal{B})\cap \pi_{m+1}$. For any $\beta \in \pi_{m+1}$ we have 
 \begin{equation}\label{beta1}
     \beta =\sum_{i=1}^m d_i \gamma_i+d_{\beta}\beta_0 =\sum_{i=1}^m \widetilde{d}_i \gamma_i+\widetilde{d}_{\beta}\widetilde{\beta}_0,
 \end{equation}    
where $d_i, \widetilde{d}_i, d_{\beta}, \widetilde{d}_{\beta} \in \mathbb{C}.$
Therefore, for any $x\in L$, we have 
 \begin{equation}\label{betas relations}
   \widetilde{d}_{\beta}= \frac{{d}_{\beta} \beta_0 (x)}{\widetilde{\beta}_0 (x)}.  
 \end{equation}
Let
 \begin{equation}\label{h form2}
  \widetilde{h}= \sum_{\pi_{m+1}} \sum_{\beta\in (\mathcal{A} \setminus \mathcal{B})\cap\pi_{m+1}}  \beta (x) \log \widetilde{d}_{\beta} \, a_{\beta}.
\end{equation}
Since by Proposition \ref{restriction_formula_h} $f$ takes values in $L$ it is sufficient to show that $\gamma(h) = \gamma(\tilde h)$ for any $\gamma \in \mathcal B$. We have
$$
\gamma(h) - \gamma(\widetilde{h}) = 
\sum_{\pi_{m+1}} \Big(\log \frac{\widetilde{\beta_0} (x)}{{\beta}_0 (x)}\sum_{\beta\in (\mathcal{A} \setminus \mathcal{B})\cap\pi_{m+1}} \gamma(a_{\beta}) \beta (x)\Big), 
$$
where the internal sum vanishes by the proof of Proposition \ref{restriction_formula_h}. Therefore, the difference $h - \widetilde h$ is a linear function with values in $L$.

\section{Identity field and flat connections}
There is a way to define an identity field for a sufficiently general solution of an oriented associativity equations corresponding to an oriented $\vee$-system $(\mathcal{A}, \Phi)$, where $\mathcal A \subset V^* \cong \mathbb C^N$. 

For an oriented $\vee$-system $(\mathcal A,\Phi)$, we call the set
\[
\mathcal V_{\mathcal A,\Phi}
=
\{\Phi(\alpha)\colon \alpha\in\mathcal A\}
\subset V
\]
the \emph{vector part} of $(\mathcal A,\Phi)$.
We say that the vector part $\mathcal V_{\mathcal A,\Phi}$ is
\emph{irreducible} if there is no non-trivial direct-sum decomposition
\[
V=V_1\oplus V_2
\]
such that 
$\Phi(\alpha)\in V_1\ \text{or}\ \Phi(\alpha)\in V_2$ for every $\alpha\in\mathcal A.$
For every $\alpha\in \mathcal{A}$, let $a_{\alpha}=\Phi(\alpha).$
The following statement takes place.
\begin{lemma}\label{lemma.T.scalar}
Suppose that the vector part $\mathcal V_{\mathcal A,\Phi}$ of an oriented $\vee$-system
$(\mathcal A,\Phi)$ is irreducible and that
it spans $V$. 
Consider the linear operator
$T\colon V\to V$
given by
$$T(v)
=
\sum_{\alpha\in\mathcal A}
\alpha(v) a_\alpha, 
\quad v\in V.$$
Then
\begin{equation}\label{T linear operator}
T=\lambda\,\text{Id}_V,    
\end{equation}
where $\lambda=\frac{1}{N} \sum_{\alpha\in\mathcal A}
\alpha(a_\alpha).$
\end{lemma}

\begin{proof}

By the oriented $\vee$-conditions \eqref{O.A. v-system.2}, for every
$\gamma\in\mathcal A$ the vector $a_{\gamma}$ is an eigenvector of
$T$ with an eigenvalue $\lambda_\gamma\in\mathbb C.$
Let 
$\Lambda
=
\left\{
\lambda_\gamma\colon \gamma\in\mathcal A
\right\}$
be the set of eigenvalues (without repetitions). 
For each $\lambda\in\Lambda$, define
\[
V_\lambda
=
\text{span}
\left\{
\Phi(\alpha)
\colon
\alpha\in\mathcal A,\ 
\lambda_\alpha=\lambda
\right\}.
\]
Since the vectors $a_\alpha$ span $V$, we get a decomposition
\[
V
=
\bigoplus_{\lambda\in\Lambda}V_\lambda.
\]
By irreducibility of $\mathcal V_{\mathcal A,\Phi}$ we have to have that the set $\Lambda$ consists of a single element $\{\lambda\}$ and therefore $T=\lambda\,\text{Id}_V.$
By taking the trace of the relation $T=\lambda\,\text{Id}_V$ we get the required value of $\lambda$. 
\end{proof}
In the next proposition we are interested in the case when conditions of Lemma \ref{lemma.T.scalar} are satisfied and $\lambda \ne 0$. We formulate the statement slightly more generally.

\begin{proposition}\label{identity-field-prop}
   Let $(\mathcal{A},\Phi)$ be an oriented $\vee$-system, and suppose that the operator
$$T=\sum_{\alpha\in \mathcal{A}}\alpha\otimes a_\alpha$$
has the form
$$T=\lambda\operatorname{Id}_V$$
for some nonzero constant $\lambda$. Then the vector field
$e=\frac{1}{\lambda}E,$
where $E$ is the Euler vector field, is the identity for the associative multiplication associated with $(\mathcal{A},\Phi)$.
\end{proposition}

\begin{proof}
For any $v\in T_x V$, $ x\in V_\Sigma$,  we have
\begin{equation*}
   e\ast v =\frac{1}{\lambda} \sum_{\alpha\in\mathcal A}
\frac{\alpha(E)\alpha(v)}{\alpha(x)}a_\alpha
=\frac{1}{\lambda}
\sum_{\alpha\in\mathcal A}
\alpha(v)a_\alpha
= \frac{1}{\lambda} T(v)=v. 
\end{equation*}
\end{proof}
\begin{example}
  Let us consider the five-parametric family of $B_2$ type oriented $\vee$-systems defined by formulas
  \eqref{B2-configuration} and  \eqref{B2-parametrization}.
Then 
\[
e=
\frac{1}{t_1+2t_3+t_4}
\left(
x_1\frac{\partial}{\partial x_1}
+
x_2\frac{\partial}{\partial x_2}
\right)\]
assuming that $\lambda=t_1+2t_3+t_4\neq 0.$
\end{example}
The associativity equations for the  multiplication \eqref{product2}  can also be reformulated in terms of flat connections as follows.
For any $\mu\in \mathbb{C}$ we define a (deformed) connection $\nabla^{\mu}$ by
\begin{equation}\label{deformed connection}
    \nabla^{\mu}_{u}=\partial_u +\mu  {\mathcal C}_u,
\end{equation}
where $u \in \Gamma\big(T_{\ast}V_ \Sigma\big)$, $\partial_u$ is the derivative in the direction of $u$, 
and the operator $\mathcal C_u\colon V\to V$ is given by the multiplication
$\mathcal C_u(v)=u*v$ for an arbitrary field $v \in \Gamma (T_*V_\Sigma)$.

The next statement is well known in greater generality for oriented associativity equations. 

\begin{proposition}\label{flatness and oriented V system}
The connection $ \nabla^{\mu}$ defined by \eqref{deformed connection} is torsion free.
Moreover, the connection is flat for any (equivalently, all) non-zero $\mu$ if and only if $(\mathcal{A}, \Phi)$ is equivalent to an oriented $\vee$-system.   
\end{proposition}
\begin{proof}

The torsion tensor of the connection $\nabla^\mu$ is
\begin{equation*}
  T^{\mu}(u,v)
=
\nabla^\mu_u v
-
\nabla^\mu_v u
-
[u,v]=\bigl(
\partial_u v-\partial_v u-[u,v]
\bigr)
+
\mu\bigl(
u\ast v-v\ast u
\bigr)=0
\end{equation*}
since the standard connection $\nabla^0$ is torsion-free,
and the multiplication $\eqref{product2}$ is commutative. 

We next consider the curvature of $\nabla^\mu$. Choose linear
coordinates $x_1,\ldots,x_N$ in $V$ and put
$\partial_i=\frac{\partial}{\partial x_i}.$
Let $\mathcal C_i$ denote the endomorphism of $T_*V$ defined by
\[
\mathcal C_i(v)=\partial_i\ast v.
\]
Thus the matrix $C_i$ of the operator $\mathcal C_i$ is 
\[
(C_i)^p_q
=
\frac{\partial^2 f^p}
{\partial x_i\partial x_q},
\]
where $f^p$ is the $p$-th coordinate of the corresponding vector potential
\begin{equation*}
     f(x)= \sum_{\alpha\in\mathcal A}
     \alpha(x)\log\alpha(x)\,a_\alpha,
    \end{equation*}
By definition, the curvature tensor is
\[
R(X,Y)Z
=
\nabla^\mu_X\nabla^\mu_YZ
-
\nabla^\mu_Y\nabla^\mu_XZ
-
\nabla^\mu_{[X,Y]}Z, 
\]
which gives for
$
\nabla^\mu_{\partial_i}
=
\partial_i+\mu C_i
$
\begin{equation*}\label{curvature.deformed.connection}
R(\partial_i,\partial_j)
=
\mu\bigl(
\partial_iC_j-\partial_jC_i
\bigr)
+
\mu^2[C_i,C_j].
\end{equation*}
The statement follows by Proposition~\ref{asossiative=oriented} and Theorem~\ref{identity=O.A.E} since $\partial_i C_j=\partial_j C_i$.
\end{proof}

\begin{remark}
Flat connections of the form \eqref{deformed connection}, \eqref{product2} (without the underlying multiplication $\ast$) were considered in \cite{dBP1}, \cite{dBP2} together with some related operations such as taking restrictions, and a description of all such connections in $\mathbb{C}^2$. Hence these results together with Theorem~\ref{identity=O.A.E}
may give another way to derive some of the above statements on the oriented $\vee$-systems. 
\end{remark}

The next result allows to relate the oriented $\vee$-systems with flat $F$ manifolds as defined e.g. in \cite{Cotti2024}.

\begin{proposition}\label{lambda for flat e}
Under the assumptions of Proposition \ref{identity-field-prop}, let $e=\frac{1}{\lambda}E$ 
be the identity vector field of the corresponding multiplication \eqref{product2}.
Then the vector field $e$ is flat with respect to  
connection \eqref{deformed connection} for $\mu = - \lambda^{-1}$.
%
\end{proposition}
\begin{proof}
Let $u$ be an arbitrary vector field. 
Then
\[
\partial_u e=\frac{1}{\lambda}u
\]
since the Euler field 
$E=\sum_{i=1}^{N}x_i\frac{\partial}{\partial x_i}$. 
Therefore we have
\begin{equation*}
    \nabla^\mu_u e=\partial_u e+\mu\,u\ast e
=\left(\frac{1}{\lambda}+\mu\right)u,
\end{equation*}
which is equal to $0$ if $\mu = - \lambda^{-1}$.
\end{proof}

\begin{corollary}
    Under the assumptions of Proposition \ref{identity-field-prop}, the corresponding associative multiplication \eqref{product2} with  the flat connection $\nabla^{-\lambda^{-1}}$ and the unity $e$  define the structure of a flat $F$-manifold on $V_\Sigma$. 
\end{corollary}
 

In the terminology of \cite{Manin2005}, Proposition \ref{flatness and oriented V system} implies that the standard connection $\nabla$, as well as connections $\nabla^\mu$, define flat structures compatible with multiplication $\ast$. 
In turn, this implies that there is an underlying $F$-manifold structure associated with any oriented $\vee$-system.
By Proposition \ref{lambda for flat e},
the choice of flat connection $\nabla^{-{\lambda}^{-1}}$ gives an $F$-manifold with a compatible flat structure including a flat identity (cf.~\cite{Manin2005}). 

\section{Oriented $\vee$-systems from complex reflection groups}

Let $G\subset GL(V)$, where $V\cong \mathbb{C}^N$, be a finite complex reflection group.
Let $\mathcal{H}$ be the corresponding arrangement of the reflection hyperplanes. 
For any $H\in \mathcal{H}$ we choose a covector $\alpha_H \in V^{\ast}$ such that $H= \{ x\in V\colon \alpha_H(x)=0 \}.$
Let $\langle , \rangle$ be a $G$-invariant non-degenerate Hermitian form on $V$. 
Let $\pi_{H}: V\to V$ be the unitary projector into the direction orthogonal to $H$ with respect to $\langle , \rangle$. 
Let us also choose $k\colon \mathcal{H}\to \mathbb{C}^\times$ which is $G$-invariant, and denote $k_{H}=k(H)$.

Define $a_{H}\in V$ such that $\pi_{H}=k_{H}^{-1}\,\alpha_H\otimes a_H$.
Define the function $f\colon V\setminus \cup_{H \in\mathcal{H}}H\to V$ by
\begin{equation}\label{f of complex reflection}
    f(x)=\sum_{H\in \mathcal{H}}\alpha_H(x) \log \alpha_H(x) a_{H}.
\end{equation}
Let $\mathcal{A}=\{\alpha_H:H\in \mathcal{H} \}$.
Let $\Phi:\mathcal{A}\longrightarrow V\setminus\{0\}$
be defined by $\Phi(\alpha_H)=a_H.$
\begin{proposition}
    The function $f(x)$ defined by \eqref{f of complex reflection} satisfies the oriented associativity conditions and the pair $(\mathcal{A}, \Phi)$ is an oriented $\vee$-system.
\end{proposition}
\begin{proof}
    The connection associated with function $f$ has the form 
\begin{equation}
    \nabla_u =\partial_u - \sum_{H\in \mathcal H}\frac{\alpha_H(u)}{\alpha_H (x)}k_{H}\,\pi_{H},
\end{equation}
which is known to be flat by \cite{CHL.2005}.
The statement follows by Proposition \ref{flatness and oriented V system} and Theorem~\ref{identity=O.A.E}.
\end{proof}
By taking subsystems and restrictions, one can obtain further oriented $\vee$-systems from complex reflection groups. 
In the case of infinite imprimitive family this leads to a {\it cyclotomic} generalization of the multi-parameter family of usual $\vee$-systems from \cite{ChalykhVeselov2001} which are related to Lauricella flat connections.  In the next subsection we start by defining the cyclotomic Lauricella configuration directly.
\subsection{Multi-parameter family of oriented $\vee$-systems}


We define a multi-parameter family of oriented $\vee$-systems in $\mathbb C^n$.
Let 
$\xi=e^{2\pi i/m}$ be the $m$th primitive root of 1 with $m \ge 2$, and let $m_1,\ldots,m_n \in \mathbb{C}.$ 
%
Consider
\begin{equation}
\label{covectors of CRG}
\mathcal A_m
=
\{\alpha_i:1\leq i\leq n\}
\cup
\{\alpha_{ij}^k:
1\leq i<j\leq n,\;
0\leq k\leq m-1\},
\end{equation}
where $\alpha_i = e^i$, $\alpha_{ij}^k = \xi^ke^i-e^j$. 
Let $B\in\mathbb C$ and let $c_i= B+m(m_i-1)$ for $1\le i \le n$.  
Define
$\Phi_m:\mathcal A_{m}\longrightarrow V$
by
\begin{equation}
\label{vector1}
    \Phi_m(\alpha_i)=
 c_i e_i
=:  a_i,
\end{equation}

\begin{equation}
\label{vector2}
\Phi_m(\alpha_{i,j}^{k})
=
\xi^{-k}m_je_i-m_ie_j=:a_{ij}^{k}.
\end{equation}
If $m_i =m_j =0$ then we assume that $\alpha_{ij}^{k}$ is not in the configuration $\mathcal A_m$. Likewise, 
if $c_i= 0$ then we assume that $\alpha_i$ is not in the configuration $\mathcal A_m$.

The following statement takes place. In the proof we assume for simplicity that $c_i, m_i \ne 0$. The proof in the general case follows the same lines. 

\begin{theorem}\label{weighted.Gmn.oriented.V}

The pair $(\mathcal A_m,\Phi_m)$ is an oriented $\vee$-system.
\end{theorem}

\begin{proof}
Let us consider a plane $\pi\subset V^*$ such that $|\pi \cap \mathcal{A}_m|\geq 2.$ 
We have the following three cases.

\medskip
\noindent
{Case 1.}
Fix $i<j$ and consider the plane
$\pi
=
\langle e^i,e^j\rangle.$
Then
\[
\pi\cap\mathcal A_{m}
=
\{e^i,e^j\}
\cup
\{\alpha_{ij}^{k}:0\leq k\leq m-1\}.
\]

First, take $\gamma=e^i$. We have
\[
e^i(a_i)=c_i,
\qquad
e^i(a_j)=0,
\qquad
e^i(a_{ij}^{k})=\xi^{-k}m_j.
\]
Hence
\[
\begin{aligned}
\sum_{\beta\in\pi_{ij}\cap\mathcal A_m}
e^i(a_\beta)\beta
&=
c_ie^i
+
m_j\sum_{k=0}^{m-1}
(e^i-\xi^{-k}e^j)
=
(c_i+mm_j)e^i
\end{aligned}
\]
using
$\sum_{k=0}^{m-1}\xi^{-k}=0$.
Thus \eqref{O.A. v-system.1} holds with
\[
\lambda_{ij}
=
B+m(m_i+m_j-1).
\]

Similarly, 
$$\sum_{\beta\in\pi_{ij}\cap\mathcal A_m}
e^j(a_\beta)\beta
=c_j e^j
-
m_i\sum_{k=0}^{m-1}
(\xi^k e^i-e^j)
=
(c_j+mm_i)e^j=\lambda_{ij} e^j.$$
By linearity, conditions \eqref{O.A. v-system.1} hold for any $\gamma \in \pi \cap \mathcal{A}_m$. 
Conditions \eqref{O.A. v-system.2} follow by Proposition \ref{Prop.pairing} since for general values of parameters $V_\pi = \langle e_i, e_j\rangle \cong \pi^*.$ 


\medskip
\noindent
{Case 2.}
Let $i,j,\ell$ be distinct indices and consider
\[
\alpha
=
\xi^ke^i-e^j,
\qquad
\beta
=
\xi^re^j-e^\ell.
\]
Consider the plane
$\pi=\langle\alpha,\beta\rangle.$
Note that this plane also contains
$\delta
=
\xi^{k+r}e^i-e^\ell.$
Thus
\[
\pi\cap\mathcal A_m
=
\{\alpha,\beta,\delta\}.
\]

The associated vectors are
\[
a_\alpha
=
\xi^{-k}m_je_i-m_ie_j,
\]
\[
a_\beta
=
\xi^{-r}m_\ell e_j-m_je_\ell,
\]
and
\[
a_\delta
=
\xi^{-(k+r)}m_\ell e_i-m_ie_\ell,
\]
which all belong to a two-dimensional plane $V_\pi$. 

We have
\[
\alpha(a_\alpha)=m_i+m_j,\qquad
\alpha(a_\beta)=-\xi^{-r}m_\ell,
\qquad
\alpha(a_\delta)=\xi^{-r}m_\ell.
\]
Hence
\[
\begin{aligned}
\sum_{\eta\in\pi\cap\mathcal A_m}
\alpha(a_\eta)\eta
&=
(m_i+m_j)\alpha
-\xi^{-r}m_\ell\beta
+\xi^{-r}m_\ell\delta\\
&=
(m_i+m_j+m_\ell)\alpha.
\end{aligned}
\]
since $\delta=\xi^r\alpha+\beta.$
Thus \eqref{O.A. v-system.1} for $\gamma=\alpha$ holds
with
\[
\lambda_{\gamma,\pi}=\lambda_\pi=m_i+m_j+m_\ell.
\]
Similarly, 

\[
\begin{aligned}
\sum_{\eta\in\pi\cap\mathcal A_m}
\beta(a_\eta)\eta
&=
(m_i+m_j+m_\ell)\beta.
\end{aligned}
\]
Thus, condition \eqref{O.A. v-system.1} holds for this case.

With respect to the bases $\alpha,\beta$ of $\pi$ and
$a_\alpha,a_\beta$ of $V_{\pi}$, the natural pairing has the matrix
\[
\begin{pmatrix}
m_i+m_j & -\xi^{-r}m_\ell\\
-\xi^rm_i & m_j+m_\ell
\end{pmatrix},
\]
whose determinant is
\[
m_j(m_i+m_j+m_\ell)\neq0.
\]
Hence the pairing is non-degenerate for general values of the parameters. Thus by Proposition \ref{Prop.pairing} the oriented $\vee$-conditions \eqref{O.A. v-system.2} follow for general and hence arbitrary values of the parameters.

\medskip
\noindent
{Case 3.}
It remains to consider a two-dimensional plane
$\pi\subset V^*$ such that $|\pi \cap \mathcal{A}_m|=2.$
We have planes $\pi$ of the forms:
\[
\left\langle e^i,\xi^ke^j-e^\ell\right\rangle,
\qquad
i,j,\ell\ \text{distinct},
\]
and
\[
\left\langle
\xi^ke^i-e^j,\,   
\xi^re^p-e^q
\right\rangle,
\qquad
\{i,j\}\cap\{p,q\}=\varnothing .
\]
Let $\pi \cap \mathcal{A}_m =\{ \alpha, \beta \}.$
Then
$\alpha(a_\beta)=\beta(a_\alpha)=0.$
Therefore conditions \eqref{O.A. v-system.1} and \eqref{O.A. v-system.2} hold for any $\gamma \in \pi \cap \mathcal{A}_m$ with $\lambda_{\gamma, \pi}=\gamma(a_{\gamma}).$
\end{proof}

One can arrive at the cyclotomic Lauricella configuration \eqref{covectors of CRG} -- \eqref{vector2} with positive integer parameters $m_i$ by taking restriction of an oriented $\vee$-system for a complex reflection group 
$G(m,1,N)$. 
Recall that its reflecting hyperplanes are
\[
H_i=\{x\in V:x_i=0\},
\qquad 1\leq i\leq N,
\]
and
\[
H_{ij}^{k}
=
\{x\in V:\xi^k x_i-x_j=0\},
\qquad
1\leq i<j\leq N,\quad 0\leq k\leq m-1.
\]

This way one starts with the configuration of covectors $\alpha_i$ $(1\le i \le N)$, $\alpha_{ij}^k$ $(1\le i<j \le N)$ in $\mathbb{C}^N$ with 
$$\Phi({e^i})=B e_i,\quad \Phi(\xi^{k}e^i-e^j )
=\xi^{-k}e_i-e_j.$$

The subspace $L$ can be taken as
\begin{align*}
L=
\begin{dcases}
x_{1}=\xi^{-k_2}x_2=\dots =\xi^{-k_{m_1}}x_{m_{1}},\\ 
x_{m_{1}+1}=\xi^{-k_{m_1 +2}}x_{m_1 +2}=\dots =\xi^{-k_{m_1 +m_2}}x_{m_{1}+m_{2}},\\
\vdots\\
x_{\sum_{i=1}^{n-1}m_i +1}=\dots =\xi^{-k_N}x_{N},
\end{dcases}
\end{align*}
where $\sum_{i=1}^{n} m_i = N$ and $k_i \in \mathbb{N}$. 
The resulting oriented $\vee$-system for any choice of $k_i$ is equivalent to the cyclotomic Lauricella configuration \eqref{covectors of CRG} -- \eqref{vector2}. 

\begin{remark}
It may be interesting to compare the oriented $\vee$-system \eqref{covectors of CRG}--\eqref{vector2} with the pullback of the Lauricella connection under the map sending every coordinate to its $m$-th power. It is pointed out in \cite{Panov2009} in the case of dimension 3 that such a pullback gives a flat connection with the singularities at the reflection arrangement of the group $G(m,1,3)$.
\end{remark}


\section*{Acknowledgments}
We are grateful to Dmitri Panov for useful discussions.
We have used ChatGPT-6 Astra for proofreading. 
The work of M.\,F.\ was partially supported by the Engineering and Physical Sciences Research Council [grant number  EP/W013053/1].

\end{document}